\newcommand{\semidirprod}{{\rm C}\hskip-11pt\times}

\newcommand{\be}{\begin{equation}}\newcommand{\ee}{\end{equation}}
\newcommand{\bea}{\begin{eqnarray}}\newcommand{\eea}{\end{eqnarray}}
\newcommand{\nn}{\nonumber}\newcommand{\p}[1]{(\ref{#1})}
\documentstyle[12pt]{article}
\begin{document}

    \thispagestyle{empty}
\begin{flushright}
JHU-TIPAC-920010\\  BONN-HE-92-07\\hepth@xxx/9203051
\end{flushright}

\begin{center} {\bf
\Large{\bf A twistor-like $D=10$ superparticle action with manifest
$N=8$ world-line supersymmetry}}\end{center} \vskip 1.0truecm

\centerline{{\bf A. Galperin}$\, ^{\dag}$}
\vskip5mm
\centerline{\it Department of Physics and Astronomy }
\centerline{\it The Johns Hopkins University}
\centerline{\it Baltimore, Maryland 21218, USA}
\vskip5mm
\centerline{\bf and}
\vskip5mm
\centerline{{\bf E.Sokatchev}$\, ^{\ddag}$}
\vskip5mm
\centerline{\it Physikalisches Institut}
\centerline{\it Universit\"at Bonn}
\centerline{\it Nussallee 12, D-5300 Bonn 1, Germany}
 \vskip 1.4truecm \bigskip   \nopagebreak

\begin{abstract}
We propose a new formulation of the $D=10$ Brink-Schwarz superparticle
 which is manifestly invariant under both the target-space
super-Poincar\'e group  and the  world-line local $N=8$
superconformal group. This twistor-like construction naturally
involves the sphere $S^8$ as a coset space of the $D=10$ Lorentz
group. The action contains only a finite set of auxiliary fields, but
they appear in unusual trilinear combinations. The origin
of the on-shell $D=10$ fermionic $\kappa$ symmetry of the standard
Brink-Schwarz formulation is explained. The coupling to a $D=10$ super-Maxwell
background requires a new mechanism, in which
the electric charge appears only on shell as an integration constant.
\end{abstract}
 \bigskip \nopagebreak \begin{flushleft} \rule{2
in}{0.03cm} \\ {\footnotesize \ ${}^{\dag}$  On leave from the
Laboratory of Theoretical Physics, Joint Institute for Nuclear
Research, Dubna, Russia}
\\ {\footnotesize \ ${}^{\ddag}$  On leave from the
Institute for Nuclear
Research and Nuclear Energy, Sofia, Bulgaria}
\vskip8mm
 March 1992
 \end{flushleft}

\newpage\setcounter{page}1

\section{Introduction}

In supersymmetric string theory there are two essentially different
approaches (see \cite{GSW} for a review).
The fundamental underlying symmetry of the Neveu, Schwarz and Ramond
approach is the world-sheet superconformal group, but there is no
supersymmetry in the target space. In the Green and Schwarz
construction it is just the other way around:  Target-space supersymmetry
 is manifest, while there is no supersymmetry on
the world sheet. Instead, in the ``magic'' dimensions $D=3,4,6$ and $10$
the GS superstring possesses  a mysterious fermionic
$\kappa$ symmetry \cite{S} of unclear geometric origin. This symmetry
is crucial for the consistency of the GS superstring. At the same
time, it has proven to be a very serious obstacle for the Lorentz covariant
quantization of the theory.

Remarkably enough, under certain conditions the above theories
are equivalent in {\it the light-cone gauge}, which means with broken
manifest Lorentz covariance. The natural question arises: If the NSR
and GS approaches are just the two faces of the same theory, is it
possible to find a new formulation that would combine the attractive
features of both? In other words, it  should be invariant under  {\it
both} the world-sheet superconformal group and target-space
Poincar\'e supersymmetry. At the same time, it should not have any
$\kappa$ invariance.\footnote{ It should be stressed that the
so-called spinning superstrings and superparticles \cite{DUR}, which
have combined target-space and world-sheet supersymmetries,
 have also $\kappa$ invariance and propagate extra degrees
of freedom, so they are {\it not} reformulations of the GS theory.}
Clearly, in such a hypothetic formulation  the light-cone gauge
linking the GS and NSR approaches should be properly  Lorentz-covariantized.

The same question can  be posed in the analogous, but simpler
problem of the Brink-Schwarz superparticle moving in a $D=3,4,6$ or
$10$-dimensional superspace.
The BS superparticle is the infinite tension limit of the  GS superstring;
being much simpler,  it still preserves  some features of the GS
theory, notably its $\kappa$ symmetry. Recently, there has been an
important development in this direction. The pioneering step was
made by Sorokin, Tkach, Volkov and Zheltukhin \cite{STV},
 who proposed
a new action for the $D=3$ superparticle with $\kappa$ symmetry replaced
by $N=1$ world-line superconformal symmetry.  The bosonic part of the
STVZ action
\be\label{1.1}
S_{\scriptsize STVZ}^{\scriptsize bose}=
\int d\tau p_\mu[\dot x^\mu - \psi\gamma^\mu\psi]
\ee
is different from the
usual massless particle action
\be\label{1.2}
S=\int d\tau [p_\mu\dot x^\mu-{1\over2}ep^2].
\ee
An essential ingredient of \p{1.1} is the
``twistor-like'' variable  $\psi$ which is a commuting spinor of the
$D=3$ Lorentz group. It is only on shell that one can show the
classical equivalence of the two theories. The
Chern-Simons nature of the STVZ action, {\it i.e.} the absense of the
world-line metric $e$ in it, was emphasized by Howe and Townsend \cite{HT}.

It is natural to try to generalize the STVZ action for the cases
$D=4,6$ and $10$, where the $D-2$ parameters of $\kappa$ symmetry
could be replaced by $N=2,4,8$ world-line supersymmetries.
Delduc and Sokatchev \cite{DS} made a step further by constructing
$D=4,6$ twistor-like superparticle actions with $N=2,4$ world-line
superconformal symmetry, respectively.\footnote{Actually, in
\cite{STV} an action for the case $N=2 \; D=4$, different from that
of \cite{DS}, was also given. The equivalence of the two actions was recently
demonstrated in \cite{Pashnev}.}  Independently, Ivanov and
Kapustnikov \cite{IK} proposed a superstring action of this type for
$N=2\; D=4$.  Like in \p{1.1}, these new actions
contain a light-like vector $v^\mu= \psi\gamma^\mu\psi$
made out the (commuting) Majorana spinor $\psi$ with 4 and 8 real
components in $D=4$ and $D=6$.

However, the attempts to generalize the STVZ approach to
the $D=10$ case have hit upon an obstacle. While $D=10$ superparticle
(or superstring) actions with $N=1$ or $2$ world-line supersymmetry
have been successfully constructed \cite{B1},
\cite{TONIN}, \cite{B2}, it seemed very hard to imagine an action
with full $N=8$ supersymmetry. The reason for this difficulty can be
explained as follows. In $D=10$, like in the
lower-dimensional cases $D=3,4,6$ there also exists a twistor-like
representation of a light-like vector $v^\mu, \;v^2=0$ in terms of a
(commuting)
Majorana-Weyl spinor $\psi^\alpha$, $\alpha=1,2,\dots,16$: $v^\mu=
\psi\gamma^\mu\psi$. The correspondence between $v^\mu$ and
$\psi^\alpha$ is, however, not one-to-one: the null vector has
10-1=9 components, while the spinor has 16. The 7 extra components
of the spinor can not be attributed to any gauge {\it  group} acting
on $\psi$,  in contrast to the cases $D=4$ and $D=6$, where the
extra components can be accounted  for as $U(1)$ and $SU(2)$ gauge
{\it groups}, respectively. Instead, the 7 gauge transformations in the
case $D=10$ form an algebra with field-dependent structure constants
\cite{B}\footnote{In fact, this is a ``physicist's derivation'' of
the famous Hopf fibrations: $S^1\rightarrow S^1$ (fibre $Z_2$),
$S^3\rightarrow S^2$ (fibre $S^1$), $S^7\rightarrow S^4$ (fibre
$S^3$) and $S^{15}\rightarrow S^8$ (fibre $S^7$). Here the l.h.s
spheres $S^1, S^3, S^7, S^{15}$ correspond to Majorana spinors in
$D=3,4,6,10$ (in $D=10$ it is a Majorana-Weyl spinor), which are
considered modulo scale transformations.  The r.h.s spheres $S^1,
S^2, S^4, S^8$ correspond to the light-like vectors in $D=3,4,6,10$
constructed in terms of the above spinors.   Finally, the fibres
are equivalents of physicist's gauge transformations. In
$D=3,4$ and $6$ the fibres are groups, since $S^1=U(1), S^3=SU(2)$.
However, in $D=10$  the fibre $S^7$ is {\it not a group}.}.
Therefore, it seemed  impossible to construct a $D=10$ STVZ-like
superparticle action with a linear realization of all the symmetries.

A way to overcome this difficulty was found by Galperin, Howe and
Stelle \cite{GHS}. They gave a group-theoretical analysis of the
above $D=3,4$ and $6$ actions and identified the light-like vectors
as the parameters of the cosets of the corresponding Lorentz groups
$SO(1,D-1)$ by their {\it maximal} subgroups. Then the  Majorana
spinors (with their gauge groups) naturally arise as the
parameters of the same cosets in the spinor representation. However,
for the case $D=10$ the GHS construction gives a different recipe how
to construct light-like vectors in terms of spinors. Instead of taking
{\it one} MW spinor $\psi^\alpha$ one should take {\it eight} MW
spinors $\psi^\alpha_a$, $a=1,\ldots,8$, which are, however, properly
{\it constrained} by the
$SO(1,9)$ conditions. This allows for {\it linear} realizations of
all the symmetries and defines the ``twistor-like'' variables
appropriate for $D=10$.

The main open problem after the work of \cite{GHS} was how to find an
action, where all those symmetries are manifest. The answer is given
in this paper. We generalize the results of \cite{STV} and \cite{GHS} and
propose an essentially new action for the most interesting case of the $D=10$
superparticle. In some sense our action is the most straightforward
generalization of the $N=1\; D=3$ STVZ superfield action to the $N=8\; D=10$
case. At first sight it might even seem too naive a
generalization to have any chances to work. Miraculously,
however, the various dangers are avoided and it does work! Moreover,
 the group-theoretical description of the $D=10$
light-like vectors in terms of spinors developed in \cite{GHS} comes
out automatically from that action. The action possesses manifest
$N=8$ world-line superconformal invariance as well as $D=10$ target-space
Poincar\'e supersymmetry. Besides the usual target-superspace coordinates
$X^\mu$ and $\Theta^\alpha$ it also involves a Lagrange multiplier with its
own abelian gauge group. The action contains a finite number  of
auxiliary fields, some of which appear in trilinear combinations.
This allows to break the ``$N=8$ barrier'' that one might anticipate
in such a theory.

We emphasize that only the classical theory of the superparticle
is discussed here. The quantum theory will be the subject of a future
publication.

The paper is organized as follows. In Section 2 we introduce the
action and discuss  its symmetries and kinematics. In Section 3  we
study the equations of motion for $X$ and $\Theta$, their
group-theoretical meaning and their solution in the light-cone gauge.
We show that the on-shell dynamics of the superparticle naturally
gives rise to a Lorentz-harmonic (twistor-like) description of the
sphere $S^8$. The large $N=8$ superfields $X$ and $\Theta$ are reduced
on-shell and in a conformal supersymmetry gauge to the usual
dynamical variables of the superparticle. In Section 4 we analyse the
equations of motion for the Lagrange multiplier and fix a Wess-Zumino
gauge for its abelian gauge invariance, in which only a single
constant vector (the particle momentum) survives. We also demonstrate
the on-shell equivalence to the BS superparticle, and discuss the
origin of $\kappa$ symmetry in the latter.  In section 5 we couple the
superparticle to a $D=10$ supergravity and super-Maxwell background.
The Maxwell coupling might seem impossible by dimensional reasons.
Its existence is due to an unusual phenomenon: the electric charge
appears in the theory only as an integration constant.  The
consistency of the coupling puts constraints on the
$D=10$ bacground. Finally, in Section 6 we compare the $D=4$ and $D=6$
analogs of our action with the results of \cite{DS} (in the case
$D=3$ our action coincides with that of STVZ \cite{STV}). It turns out
that in our formulation the complex structures of the $D=4,6$ theories,
which were put in the basis of the constuction of \cite{DS}, manifest
themselves only on shell. We conclude the paper by formulating
further problems that may be solved along the lines developed here.

\section{The  action and its invariances}

We propose the following action for a superparticle moving in D=10
Minkowski superspace:
\be\label{2.1}
S=\int d\tau d^8\eta P_{a\mu}(D_aX^\mu -
iD_a\Theta\gamma^\mu\Theta).
\ee
It is an integral over the $N=8$ world-line superspace
$(\tau, \;\eta_a) \;\;(a=1,2,\ldots,8)$, which contains one even and eight
odd coordinates. $D_a$ are the $N=8$ supercovariant derivatives
\be\label{2.2}
D_a=\partial_a + i\eta_a\partial_\tau\;,\;\;
\{D_a,\;D_b\}=2i\delta_{ab}\partial_\tau \;\;\;\;\;
(\partial_a=\partial/\partial\eta^a).
\ee
Note the natural $SO^\uparrow(1,1)\times O(8)$ automorphism of the
$D$ algebra.
The dynamical variables  $P(\tau,\eta)$, $X(\tau,\eta)$ and
$\Theta(\tau,\eta)$ are $N=8$ superfields. The anticommuting
superfield $P_{a\mu}$ (which we shall call a Lagrange multiplier)
carries a world-line $O(8)$ index $a$ and a ten-dimensional Lorentz
vector index $\mu$. $X^\mu$ and $\Theta^\alpha$ are the coordinates
of $N=1 \; D=10$ target superspace, $\Theta^\alpha$ is a
Majorana-Weyl spinor, which has 16 real components in $D=10$.

The action \p{2.1} has a number of symmetries. First of all, it is
manifestly invariant under $N=1\; D=10$ ${\it rigid}$ target-space
supersymmetry:
\be\label{2.3}
\delta\Theta^\alpha=\epsilon^\alpha,
 \;\;\delta X^\mu=i\Theta\gamma^\mu\epsilon, \;\;\delta P_{a\mu}=0
\ee
and under the orthocronous ten-dimensional Lorentz group $O^\uparrow(1,9)$.

The action also has two types of ${\it gauge}$ invariances. The  first one is
the $N=8$ world-line superconformal group which can be defined as the
subgroup of the general diffeomorphism group
$\tau\rightarrow\tau'(\tau,\eta), \;
\eta\rightarrow\eta'(\tau,\eta)$ that transforms the flat world-line
spinor derivatives $D_a$ \p{2.2} homogeneously. The defining constraint and the
transformation law are
\be\label{2.4}
  D'_a\tau-i\eta_bD'_a\eta_b=0 \;\;\Rightarrow \;\;
D'_a=(D'_a\eta_b)D_b.
\ee
For infinitesimal transformations the constraint in \p{2.4} can be  solved
in terms of an unconstrained world-line superfunction $\Lambda(\tau,\eta)$:
\be\label{2.5}
\delta\tau\equiv\tau'-\tau=\Lambda-{1\over 2}\eta_aD_a\Lambda, \;\;\;\;
\delta\eta_a\equiv\eta'_a-\eta_a=-{i\over 2}D_a\Lambda.
\ee
Then the superconformal transformation law of the covariant derivatives can be
rewritten in the following suggestive form
\be\label{2.6}
\delta D_a\equiv D'_a-D_a=-{1\over 2}\partial_\tau\Lambda \;D_a+
{i\over 4}[D_a,D_b]\Lambda\; D_b,
\ee
where the first term corresponds to scale transformations and the second to
$O(8)$ ones. The Weyl parameter $\partial_\tau\Lambda$ appears also in
the world-line supervolume transformations
\be\label{2.7}
\delta(d\tau d^8\eta)=-3\;\partial_\tau\Lambda(d\tau d^8\eta).
\ee
The ten-dimensional coordinates $X^\mu$ and  $\Theta^\alpha$ are
world-line scalars
\be\label{2.8}
X'(\tau',\eta')=X(\tau,\eta), \;\;\;\Theta'(\tau',\eta')=
\Theta(\tau,\eta).
\ee
The superconformal invariance of the action \p{2.1} is achieved by
the following transformation of the Lagrange multiplier:
\be\label{2.9}
\delta P^\mu_a = {7\over 2}\partial_\tau\Lambda \;P^\mu_a
+{i\over 4}[D_a,D_b]\Lambda \; P^\mu_b,
\ee
which compensates for the derivative and volume transformations
\p{2.6} and \p{2.7}.

Secondly, the Lagrange multiplier $P$ has a
large {\it abelian gauge invariance}:
\be\label{2.10}
\delta P^{\mu}_a = D_b(\xi_{abc}\gamma^\mu D_c\Theta),
\ee
where the parameter $\xi^\alpha_{abc}(\tau,\eta)$ is a
(commuting) Majorana-Weyl spinor
and it is totally symmetric and traceless with respect to the $O(8)$ indices
$a,b,c$. To check this invariance
one integrates $D_b$ by parts and uses the $O(8)$ properties of
$\xi_{abc}$ and the ten-dimensional $\gamma$-matrix identity
\be\label{id}
(\gamma^\mu)_{(\alpha\beta}(\gamma_\mu)_{\gamma)\delta} =0.
\ee
As we shall see in what follows, this abelian gauge invariance of $P$
is crucial for the consistency of the action. Since it relies on the
$\gamma$-matrix identity \p{id}, which is valid in 3,4,6 and 10
space-time dimensions, it
is clear that superparticle actions similar to \p{2.1} can be written
down in all those special dimensions. In fact, in the case $D=3$ the
action \p{2.1} is just the action of STVZ \cite{STV}. Note, however,
that the case $D=3$ is exceptional, because there one has $N=1$
world-line supersymmetry  and the
right-hand side of \p{2.10} automatically vainishes. The cases $D=4$ and
$6$  will be discussed in section 6.

It should be emphasized that all the above symmetries -
the rigid $D=10$ superPoncar\'e group, the local  $N=8$
superconformal group and the Lagrange multiplier invariance, are
off-shell, independent of each other and linearly realized.

To complete the definition of the action one should postulate  the
following kinematical properties: the vector
$\partial_\tau X^\mu$ is strictly non-vanishing and the matrix
$D_a\Theta^\alpha$ is of the maximal rank, {\it i.e.}
\be\label{2.11}
\partial_\tau X^\mu \neq 0, \;\; {\rm rank}\;\vert\vert
D_a\Theta^\alpha\vert\vert=8.
\ee
Geometrically, this means that the superparticle trajectory
$X^\mu=X^\mu(\tau,\eta), \;\; \Theta^\alpha = \Theta^\alpha(\tau,\eta)$
is a non-degenerate (1,8) surface in  ten-dimensional Minkowski
superspace. At the same time, it means that the superconformal group
\p{2.5} is spontaneously broken, {\it i.e.} it can be compensated by
the superfields $X$ and $\Theta$.  This requirement will also make
possible choosing a light-cone gauge, where one component of $X^\mu$
and eight components of $\Theta^\alpha$ will be identified with
$\tau, \eta_a$ (see the next section). Besides \p{2.11}, one should
require that  the following tensor of the gauge groups
\p{2.9} and \p{2.10}
\be\label{2.12}
p^\mu \equiv\epsilon^{a_1\ldots a_8}
D_{a_1}\dots D_{a_7}P^\mu_{a_8}
\ee
do not vanish. As we shall see, this $D=10$ vector plays the r\^ole of the
superparticle momentum, which should be strictly non-vanishing in
Chern-Simons-type actions like \p{1.1} and \p{2.1}.

In sections 3 and 4 we will prove the classical on-shell equivalence
of the action \p{2.1} to the Brink-Schwarz action.

\section{Geometry of the  superparticle motion}

In this section we shall study the superparticle equation of motion
following from the variation of the action
\p{2.1} with respect to the Lagrange multiplier $P^\mu_a$:
\be\label{3.1}
D_aX^\mu-iD_a\Theta\gamma^\mu\Theta = 0.
\ee
We shall show that this equation, together with the superconfromal
group \p{2.5} allows one to express all the components of the $N=8$
superfields $X^\mu(\tau,\eta)$ and $\Theta^\alpha(\tau,\eta)$ in
terms of their lowest-order components $x^\mu(\tau)=X^\mu\vert_{\eta=0},\;
\theta^\alpha(\tau) =
\Theta^\alpha\vert_{\eta=0}$, which are the usual dynamical variables
of the superparticle.\footnote{This phenomenon has already been
observed by Berkovits \cite{B}. He proposed an equation similar to \p{3.1}
for describing part of the on-shell dynamics of the heterotic
superstring. He studied that equation in an $O(8)$ non-covariant and
highly non-linear gauge (using a hypercomplex-number notation) and
reached a similar conclusion about its content.} We shall also exhibit the
geometric meaning of the motion of the  superparticle on
an $N=8$ super-world-line. In particular, we shall emphasize the double
r\^ole of the commuting spinors $\psi^\alpha_a =
D_a\Theta^\alpha \vert_{\eta=0}$ as superpartners of $\theta^\alpha$
under $N=8$ world-line conformal supersymmetry, on the one
hand, and as twistor-like variables parametrizing the sphere $S^8$ on
the other hand.

We begin by deriving an important consequence of
eq.\p{3.1}.\footnote{Note that in an interesting recent paper
\cite{TONIN} examined an equation like \p{3.1} for the $D=10$
superstring, but with $N=2$ world-sheet supersymmetry only. From it he
derived a corollary similar to \p{3.2} and identified it as the
defining condition for ``pure spinors'' \cite{PURE}. At the same
time, in \cite{GHS} it was shown that pure spinors are a particular
case of the $D=10$ Lorentz-harmonic variables discussed below.} It is
obtained by hitting it by another spinor derivative $D_b$ and taking
 the symmetric traceless part in $a,b$:
\be\label{3.2}
D_a\Theta\gamma^\mu D_b\Theta = {1\over 8}\delta_{ab} D_c\Theta
\gamma^\mu D_c\Theta .
\ee
There is a remarkable analogy between  \p{3.1} and \p{3.2}  and the {\it
finite}
$N=8$ superconformal transformations. Hitting  the defining constraint
\p{2.4} by $D_c$ and taking the symmetric and traceless part in $a,c$,
one finds
\be\label{3.3}
D'_a\eta_b\;D'_c\eta_b={1\over 8}\delta_{ac}
D'_d\eta_b\;D'_d\eta_b.
\ee
This means that the matrix
$D'_a\eta_b$ takes its values in $SO^\uparrow(1,1)\times O(8)$, where
$SO^\uparrow(1,1)$ is represented by the positive root
$\;\lambda=(D'_d\eta_b\;D'_d\eta_b/8)^{1/2}$ and $O(8)$ by
$D'_a\eta_b/\lambda$.

There is a clear similarity between  \p{3.1}, \p{3.2} and \p{2.4},
\p{3.3}. In fact, equations \p{3.2} and \p{3.3}, together with the
non-degeneracy condition \p{2.11} and the superconformal
transformation law (see \p{2.4}, \p{2.8})
\be\label{3.4}
D'_a\Theta'^\alpha=D'_a\eta_b \;D_b\Theta^\alpha
\ee
have an important geometric meaning.
It turns out that the rank 8 matrix $D_a\Theta^\alpha$,
constrained by \p{3.2} and considered modulo the $SO^\uparrow(1,1)\times O(8)$
gauge transformations \p{3.3}, \p{3.4} parametrizes the sphere $S^8$.
 This is not surprising, since the light-like velocity $\dot x^\mu$ of
a massless $D=10$ particle does describe an eight-sphere. Indeed,
$\dot x^\mu\dot x_\mu  = 0 \ \rightarrow \ (\dot x^1)^2 + \ldots +
(\dot x^9)^2 = (\dot x^0)^2 $. This, together with
the physical assumption $\dot x^0 \neq 0$ and up to
$SO^\uparrow(1,1)$ (scale) transformations, is the definition of
$S^8$. To be more precise, in our case the eight-sphere emerges as
the following coset space of the $D=10$ Lorentz group\footnote{ In
fact, equation \p{3.2} and the gauge transformations \p{3.3} have
been proposed in \cite{GHS} in the context of the above coset space
construction. However, it was not realized there that
\p{3.2} is {\it the only} algebraic restriction on $D_a\Theta^\alpha$.}
\be\label{3.5}
S^8={{\rm Pin}(1,9)\over [SO^\uparrow(1,1)\times O(8)]\semidirprod
K^i} .
\ee
Here ${\rm Pin}(1,9)$ is the double covering group for the orthocronous
Lorentz group $O^\uparrow(1,9)$ (analogously, ${\rm Spin}(1,9)$
covers the proper orthocronous group $SO^\uparrow(1,9)$).
The denominator group  is its {\it maximal} subgroup, $K^i$ are
the conformal boost generators (in fact, $D_a\Theta^\alpha$ is
that half of the ${\rm Pin}(1,9)$  $16\times 16$ matrix which
transforms only under $SO^\uparrow(1,1)\times O(8)$ and is $K^i$
inert, see \cite{GHS} for details).

To see why the matrix $D_a\Theta^\alpha$, subject to the algebraic
constraint \p{3.2} and considered modulo the transformations \p{3.4},
corresponds to the sphere $S^8$,
it is convenient to use light-cone coordinates. Given a
D=10 vector $v^\mu= (v^0, v^i, v^9)$ one defines $v^{\pm\pm}=v^0\pm
v^9\; \rightarrow v^\mu v_\mu = -v^{++}v^{--} +(v^i)^2$; a
Majorana-Weyl spinor $\Theta^\alpha$ is decomposed into a pair
$(\Theta^-_A, \;\Theta^+_{\dot A})$, where $A$ and ${\dot A}$ are
indices of the $8_s$ and $8_c$ representations of the $O(8)$ subgroup
of the Lorentz group and the weights $\pm$ correspond to the
$SO^\uparrow(1,1)$ subgroup (in what follows we shall often omit the
$SO^\uparrow(1,1)$ weights, if this will not cause confusion). Using
a suitable representation for the $\gamma$ matrices,\footnote{We use
the following conventions: $\eta_{\mu\nu}={\rm diag}(-1,1, \ldots,
1)$,
\bea
\gamma^0=
        \left(\begin{array}{cc}
                 1 & 0 \\
                 0 & 1 \end{array}\right), \ \
\gamma^9=
        \left(\begin{array}{cc}
                 -1 & 0\\
                  0& 1 \end{array}\right), \ \
\gamma^i=
        \left(\begin{array}{cc}
                 0 & \gamma^i \\
                 (\gamma^i)^T & 0 \end{array}\right) \nn
\eea
where $\gamma^i_{A{\dot B}}$ are $8\times 8$ matrices of $O(8)$,
the indices $i, \;A$ and ${\dot B}$ correspond to the representations
$8_v, \;8_s$ and $8_c$
 of $O(8)$, respectively.} we can write down
the three light-cone projections
of \p{3.2} as follows:
\be\label{3.13}
D_a\Theta_A \;D_b\Theta_A = {1\over 8} \delta_{ab}
D_c\Theta_A\;D_c\Theta_A,
\ee
\be\label{3.14}
D_a\Theta_{A}\gamma^i_{A\dot B}D_b\Theta_{\dot B} +
(a\leftrightarrow b)=
{1\over 4} \delta_{ab}
D_c\Theta_{A}(\gamma^i)_{A\dot B} D_c\Theta_{\dot B},
\ee
\be\label{3.15}
D_a\Theta_{\dot A} D_b\Theta_{\dot A}={1\over 8} \delta_{ab}
D_c\Theta_{\dot A}\;D_c\Theta_{\dot A} .
\ee

The non-degeneracy of $D_a\Theta^\alpha$ \p{2.11}
implies that the right-hand sides of \p{3.13}
and \p{3.15} cannot vanish simultaneously. This corresponds to the two charts
 needed to cover $S^8$: on the first chart
\be\label{3.16}
D_c\Theta_A D_c\Theta_A \neq 0,
\ee
on the other chart
\be\label{3.17}
 D_c\Theta_{\dot A} D_c\Theta_{\dot A} \neq 0.
\ee
Suppose that we deal with the first chart \p{3.16} and consider the
superconformal transformation \p{3.4}
\be\label{3.18}
D'_a\Theta'_A=D'_a\eta_b \;D_b\Theta_A.
\ee
Clearly, the parameter $D'_a\eta_b$ has the same content as
$D_a\Theta_A$, see \p{3.3} (this means that the superconformal group is
spontaneously broken). The conclusion is that on the first chart
the matrix $D_a\Theta_A$ can be gauged into the unit matrix
\be\label{3.19}
D_a\Theta_A=\delta_{aA}.
\ee
This gauge identifies  the $O(8)$ and
 $SO^\uparrow(1,1)$  subgroups of the superconformal group with
the $O(8)$ and
$SO^\uparrow(1,1)$
subgroups of the Lorentz group $O^\uparrow(1,9)$. In particular, the indices
$a,b,\ldots$ now correspond  to the $8_s$ representaion of $O(8)$
and the world-line coordinates and derivatives carry the following
$SO^\uparrow(1,1)$ weights: $\tau \rightarrow \tau^{--}, \;
\eta_a \rightarrow \eta^-_a, D_a \rightarrow D_a^+, \partial_\tau
\rightarrow\partial^{++}_\tau$.

Next, we
substitute the gauge \p{3.19} in the transverse projection \p{3.14}:
\be\label{3.20}
\gamma^i_{a\dot B}D_b\Theta_{\dot B} +
(a\leftrightarrow b)=
{1\over 4} \delta_{ab}
\gamma^i_{c\dot B} D_c\Theta_{\dot B}.
\ee
The matrix $D_a\Theta_{\dot A}$ describes a reducible $O(8)$
representation,\footnote{ A useful list of the $O(8)$ representations and their
products can be found in \cite{SL}.}
$8_s\times 8_c = 8_v + 56_v$ or $D_a\Theta_{\dot
A}=\gamma^i_{a\dot A}Y^i +
\gamma^{[ijk]}_{a \dot A}Y^{ijk}$. Only the $8_v$ part solves
\p{3.20}, so we find
\be\label{3.21}
D_a\Theta_{\dot A}=\gamma^i_{a\dot A}Y^i .
\ee
Inserting this into \p{3.15} we see that it is solved as
well.

Thus, on the first chart the general
solution to \p{3.13} - \p{3.15} (modulo gauge transformations)
is parametrized by eight  coordinates $Y^i$. A similar analysis for the
second chart \p{3.17} leads to another set of coordinates
\be\label{3.22}
D_a\Theta_A=\gamma^i_{aA}Z^i, \;\;\;D_a\Theta_{\dot A}=\delta_{a\dot A}.
\ee
Now we should relate these two sets of coordinates in the overlapping
area, {\it i.e.} when both conditions \p{3.16} and \p{3.17} are satisfied.
To this end  consider the vector appearing in the r.h.s. of \p{3.2},
\be\label{3.23}
v^\mu={1\over 8}D_a\Theta\gamma^\mu D_a\Theta.
\ee
This is a light-like vector, $v^\mu v_\mu=0$ due to the identity \p{id}
and eq.\p{3.2}.
It is $O(8)$ invariant and has a nonvanishing $SO^\uparrow(1,1)$ weight.
Hence the ratio of any two of its components is gauge invariant. The two charts
described above correspond to $v^{--}\neq 0$ and $v^{++}\neq 0$,
respectively. Using \p{3.19}, \p{3.21}-\p{3.23} one finds
\be\label{3.24}
Y^i={v^i\over v^{--}}, \;\;\;Z^i={v^i\over v^{++}}.
\ee
In the overlapping area $Y^i=Z^i/(Z^j)^2$, so $Y^i$ and $Z^i$ can be
considered as stereographic coordinates of the sphere $S^8$.

In conclusion we can say that the 8 commuting spinors
(``twistor variables")  $D_a\Theta^{\alpha}\vert_{\eta=0}$ have a
double r\^ole. On the one hand, they are the superpartners of the
Grassmann coordinates of target superspace, $\theta^\alpha =
\Theta^\alpha\vert_{\eta=0}$ with respect to $N=8$ world-line
supersymmetry. On the other hand, they are Lorentz-harmonic
coordinates on the sphere $S^8$, regarded as the coset \p{3.5}. The
requirement of double supersymmetry (world-line and target-space)
establishes a natural link between these two concepts \cite{STV}, \cite{IK},
\cite{DS}, \cite{GHS}. We stress also that the sphere $S^8$ and
the related light-like vector \p{3.23} appear on shell only,
while off shell the eight Majorana spinors  $D_a\Theta^{\alpha}\vert_{\eta=0}$
parametrize a bigger manifold.

Having clarified the geometric meaning of the superparticle
equation of motion
\p{3.1} and of its gauge group \p{3.3}, \p{3.4}, now we come back to
the analysis of the content of that equation. In what follows we
shall deal with the first chart on $S^8$, so the gauge \p{3.19} is
implied below. This gauge completely fixes the superconformal group,
up to a constant translation and a supertranslation
\be\label{3.25}
\delta\tau = \rho +i\eta_a\epsilon_a, \;\; \delta\eta_a=\epsilon_a .
\ee
The general solution to \p{3.19} is given by
\be\label{3.27}
\Theta_A=\eta_A,
\ee
where we have used the constant world supersymmetry parameter
\p{3.25} to fix the possible constant term in \p{3.27}, thus
identifying the rigid $N=8$ world-line supersymmetry with one half of
ten-dimensional supersymmetry.

The above results allow us to considerably simplify the original
equation \p{3.1}. In the light-cone gauge it has the following three
projections:
\be\label{3.28}
D_aX^{--} - 2i\eta_a =0,
\ee
\be\label{3.29}
D_aX^i - i\gamma^i_{a\dot B}\Theta_{\dot B} -
iD_a\Theta_{\dot B}\gamma^i_{\dot BB}\eta_B =0,
\ee
\be\label{3.30}
D_aX^{++}-i(D_a\Theta_{\dot B})\Theta_{\dot B} =0.
\ee
The first one, eq.\p{3.28}, has the obvious solution
\be\label{3.31}
X^{--} =2 \tau^{--},
\ee
where we have fixed the possible constant term in the r.h.s. by means
of the translation from \p{3.25}.
Hitting \p{3.29} with $D_a$ and using \p{3.19}, \p{3.21} one obtains
\be
Y^{i} = {1\over 2}(\dot X^i-i\dot\Theta\gamma^i\eta),
\label{3.33}\ee
where the dot denotes $\partial/\partial\tau^{--}$.
Combining \p{3.21}, \p{3.29} and \p{3.33} one finds
\be\label{3.34}
D_a\Theta_{\dot B} ={1\over 2} \gamma^i_{a\dot B}
({\dot X}^i - i\dot\Theta\gamma^i\eta),
\ee
\be
D_aX^i=i\gamma^i_{a\dot B}\Theta_{\dot B} +{i\over 2}\gamma^k_{a\dot B}
\gamma^i_{b \dot B}\eta_b({\dot X}^k - i\dot\Theta\gamma^k\eta).
\label{3.35}\ee
This allows us to conclude that all the components of $X^i$ and
$\Theta_{\dot A}$ are expressed in terms of their lowest-order
components $x^i(\tau)=X^i\vert_{\eta=0}$ and $\theta_{\dot
A}(\tau)=\Theta_{\dot A}\vert_{\eta=0}$. It is clear from
\p{3.30} that the same applies to $X^{++}$ (the solution to \p{3.30}
will be given later, see \p{4.24}). It is rather remarkable that the
content of equation \p{3.1} is so simple, given the original
complexity of the $N=8$ superfields under consideration.

The careful reader may have noticed that \p{3.1} does not
restrict the $\tau$ dependence of the world-line fields $x^i(\tau)$ and
$\theta_{\dot A}(\tau)$. The on-shell equations for these fields will
be found in the next section.

\section{Equations of motion and gauge fixing for the Lagrange multiplier}

In the previous section we studied the equation of motion \p{3.1}
obtained from the action \p{2.1}  by varying with respect to the
Lagrange multiplier $P^\mu_a$. Usually, employing a Lagrange
multiplier is not a safe trick,
since besides the desired propagating modes in $X$ and $\Theta$,
it could also give rise to a number of extra ones, coming from the
equations for the Lagrange multiplier itself. One of the most
unexpected features of the action \p{2.1} is that this danger is
miraculously avoided. Indeed, now we shall demonstrate that the
equations of motion for the Lagrange multiplier $P^\mu_a$ following
from the variation of $X^\mu$ and $\Theta^\alpha$, together with the
powerful abelian gauge invariance \p{2.10} reduce $P^\mu_a$ to a
single constant $D=10$ vector $p^\mu$ (see \p{2.12}), which is just
the on-shell momentum of the superparticle. They also imply the
correct dynamical equations for $x^i(\tau)$ and $\theta_{\dot A}(\tau)$.

The equations under consideration are
\be\label{4.1}
D_aP^\mu_a = 0,
\ee
\be\label{4.2}
(\gamma_\mu D_a\Theta)_\alpha P^{\mu}_a =0.
\ee
Due to the complexity of the $N=8$ superfields the analysis of eqs. \p{4.1}
and \p{4.2} and of the gauge transformation \p{2.10} is rather
involved and it is worthwhile to first sketch the procedure. We
decompose these equations in the light-cone frame and use the
light-cone gauge \p{3.19}.  Then we study the $(--)$ projections of
\p{4.1} and \p{2.12} to show that $P_a^{--}$ is reduced to a single
constant component in a suitable Wess-Zumino gauge for the invariance
\p{2.10}. Next we consider the transverse projections of \p{4.1} and
\p{2.12} and the $\alpha=\dot A$ projection of \p{4.2} to find the WZ
gauge of $P^i_a$ on shell. Finally, we take the $(++)$ projections of
\p{4.1}, \p{2.12} and the $\alpha=A$ projection of \p{4.2} to find an
expression for $P_a^{++}$.

Before proceeding further, we mention that as a consequence of \p{4.1}
the invaraint of the
abelian gauge transformations $p^\mu=\epsilon^{a_1\ldots a_8}
D_{a_1}\dots D_{a_7}P^\mu_{a_8}$ \p{2.12} is a constant,
\be\label{4.2a}
D_ap^\mu=0 \; \rightarrow \; p^\mu={\rm const}.
\ee
This follows from the relation
\be\label{4.2c}
D_ap^\mu=
\epsilon_{ab_1\ldots b_7} D_{b_1}\dots D_{b_7}D_cP_c^\mu,
\ee
which can be proved using the  identities
\bea\label{4.2d}
D_{a_1}D_{[a_2}\ldots D_{a_k]}=D_{[a_1}\ldots D_{a_k]}
+i(k-1)\delta_{a_1[a_2}D_{a_3}\ldots D_{a_k]}\partial_\tau,\nn \\
D_{[a_2}\ldots D_{a_k]}D_{a_1}=D_{[a_2}\ldots D_{a_k}D_{a_1]}
+i(k-1) D_{[a_2}\ldots D_{a_{k-1}}\delta_{a_k]a_1} \partial_\tau.
\eea

Let us first study the $(--)$ light-cone projection of \p{4.1}.
Using the light-cone gauge condition \p{3.19} one finds
the following gauge transformation \p{2.10} for $P^{--}_a$:
\be\label{4.3}
\delta P^{--}_a=D_b(\xi^{D}_{abc}D_c\Theta_D)
\equiv D_b\xi_{ab},
\ee
where $\xi_{ab}=\xi^c_{abc}$
is an arbitrary symmetric traceless superfield.
On shell, the {\it only} invariant of the  gauge transformation \p{4.3}
is the  $p^{--}$ component of $p^\mu$.
This can be proved as follows. The component content of an $N=8$
worldline  superfield can be extracted by the derivatives
 $D_a, \; D_{[a_1}
D_{a_2]},\;\ldots, D_{[a_1}D_{a_2}\ldots D_{a_8]}$
at $\eta=0$. Now, the lowest-order component $P^{--}_a$
is shifted by an arbitrary
parameter $D_b\xi_{ab}$ and can therefore be completely gauged away.
The next component, $D_cP_a^{--}$ is shifted by  $D_cD_b\xi_{ab}$, which is an
arbitrary traceless parameter and hence only the trace $D_aP_a^{--}$
survives, but the latter vanishes on shell \p{4.1}.
Proceeding in this manner we find that up to the level of six derivatives
the only gauge invariant components are the derivatives of the tensor
$D_aP_a^{--}$, {\it i.e.}
 $D_aP_a^{--}$,  $D_bD_aP_a^{--}$,  $D_{[b_1}D_{b_2]}D_aP_a^{--},
\ldots$, $D_{[b_1}\dots D_{b_5]}D_aP_a^{--}$, but all of them  vanish
on shell \p{4.1}.  At the level of seven derivatives the component
$D_{[b_1}\dots D_{b_7]}P_a$ contains the $O(8)$ representations
$8_s\times 8_s= 1_s+28+35_s$, while the parameter $D_{[b_1}\dots
D_{b_7]}D_b\xi_{ab}$ contains $35_s$  only. The surviving
representaions $1_s$ and $28$ correspond to $p^{--}$ and
$D_{[b_1}\dots D_{b_6]}D_aP_a^{--}=0$, respectively. So, we
conclude that on shell and in the  $\xi$ Wess-Zumino gauge described
above the superfield $P^{--}_a$ contains one constant component only
\be\label{4.5}
P^{--}_a={1\over 7!8!}\epsilon_{ab_1\dots b_7}\eta_{b_1}\ldots
\eta_{b_7}p^{--}, \;\;p^{--}={\rm const}.
\ee

For the transverse projection $P^i_a$ the gauge transformation \p{2.10}
can be written in a way similar to \p{4.3}:
\be\label{4.6}
\delta P^i_a=D_b\xi^i_{ab},
\ee
where
\be\label{4.7}
\xi^i_{ab}=(\xi^i_{ab})_{\rm add} +
\xi^A_{abc}\gamma^i_{A\dot B}D_c\Theta_{\dot B},\;\;\;
(\xi^i_{ab})_{\rm add}\equiv\xi^{\dot B}_{abc}\gamma^i_{c\dot B}.
\ee
 Note that  $\xi^i_{ab}$ consists of two
terms, of which only the first one is field-independent, {\it i.e.}
can be used to gauge away parts of $P^i_a$.
This additive term is constrained as follows:
\be\label{4.8}
\gamma^i_{a\dot A}(\xi^i_{ab})_{\rm add}=
\delta_{\dot A\dot B}\xi^{\dot B}_{aba}=0.
\ee
It is easy to see that
this  is the only condition on the additive part.
Indeed, the parameter $\xi^i_{ab}$ is symmetric and traceless with respect to
the $O(8)$ indices $a,b$, therefore it contains $35_s\times 8_v
=56_v+224_{sv}$. At the same time,  the l.h.s. of \p{4.8}
contains $8_s\times 8_c=8_v+56_v$, so \p{4.8} kills the $56_v$
in $(\xi^i_{ab})_{\rm add}$. Thus the latter corresponds to the
irreducible representation $224_{sv}$, which cannot be  restricted any
more.

The restriction \p{4.8} implies  the existence  of another ``tensor''
\be\label{4.9}
 T_{\dot A}=\gamma^i_{a\dot A}P^i_a.
\ee
Though non-invariant under the second, field-dependent term in \p{4.7},
$ T_{\dot A}$
 is invariant under the $(\xi^i_{ab})_{\rm add}$ transformations, so it
cannot be gauged away. Instead, it is expressed in terms of the other
fields by means of the second equation for the
Lagrange multiplier \p{4.2}.  In the light-cone frame the latter reads
\be\label{4.10}
-{1\over 2}(\gamma^{++}D_a\Theta)_\alpha P^{--}_a
-{1\over 2}(\gamma^{--}D_a\Theta)_\alpha P^{++}_a
+(\gamma^i D_a\Theta)_\alpha P^i_a=0.
\ee
The $\alpha={\dot A}$ projection of this equation is given by
\be\label{4.11}
\gamma^i_{a\dot A}(P^i_a-Y^iP_a^{--})=0\ \ \rightarrow \ \ {T}_{\dot
A}=\gamma^i_{a\dot A}Y^iP_a^{--}.
\ee

Studying the WZ gauge  for $P^i_a$, one finds that up to the sixth
order all the derivatives $D_{[a_1}\ldots D_{a_k]}P^i_b, \; (k\leq
6)$ are expressed in terms of derivatives of $D_aP^i_a$ and $T_{\dot
A}$ and therefore vanish on shell and in the WZ gauge \p{4.5} for
$P^{--}_a$. The seventh order is less trivial and we give some details
for the inquiring reader. Firstly, the additive part of the gauge
parameter at this order is $D_{[a_1}\ldots D_{a_7]}D_b\xi^i_{ab}$ and
it contains $8_s\times 56_c = 8_v+56_v+160_v+224_{sv}$ (with the
restriction \p{4.8} taken into account). However, among the latter
$28\times 8_v = 8_v+56_v+160_v$ vanish due to the relation
$D_{[a_1}\ldots D_{a_6]}D_aD_b\xi^i_{ab}=0$. Hence the seventh-order
parameter contains $224_{sv}$ only. Secondly, the
seventh order in $P^i_a$, {\it i.e.} $D_{[a_1}\ldots D_{a_7]}P^i_b$
contains $8_s\times 8_s\times 8_v = (8_v)^2 + (56_v)^2+
160_v+224_{sv}$, where the $224_{sv}$ is a pure gauge. Thirdly, at
the seventh order there are the following tensor components:
$D_{[a_1}\ldots D_{a_6]}D_bP_b^i \ \rightarrow\  28\times 8_v=8_v+56_v+160_v$,
$p^i \ \rightarrow\  8_v$, $D_{[a_1}\ldots D_{a_7]} T_{\dot A} \
\rightarrow\  8_s\times 8_c=8_v+56_v$. This matches the
gauge-invariant content of $D_{[a_1}\ldots D_{a_7]}P^i_b$ except for
of one extra $8_v$. Hence there should exist a relation among the
three $8_v$ in $D_{[a_1}\ldots D_{a_6]}D_bP_b^i$, $p^i$ and
$D_{[a_1}\ldots D_{a_7]}T_{\dot A}$. It is given by
\be\label{4.13}
p^i=\gamma^i_{b\dot A}\epsilon_{a_1\ldots a_7b}D_{[a_1}\ldots
D_{a_7]}T_{\dot A} +
{7\over 2}\gamma^i_{b\dot A}\gamma^j_{c \dot A}\epsilon_{bca_1\dots a_6}
D_{[a_1}\ldots D_{a_6]}D_dP_d^j,
\ee
up to some fifth-order terms, which were already shown to vanish on shell
and in the WZ gauge. Using \p{4.1} and \p{4.11} we find
\be\label{4.14}
p^i=\gamma^i_{b\dot A}\epsilon_{a_1\ldots a_7b}D_{[a_1}\ldots
D_{a_7]}(\gamma^k_{c\dot A}Y^kP^{--}_c).
\ee
Substituting \p{3.33} and \p{4.5} in \p{4.14}, we arrive at
the important equations
\be\label{4.15}
p^i={1\over 2}{\dot x}^i p^{--}, \;\;p^{--}{\dot\theta}_{\dot A}=0.
\ee
which will be discussed later on, see \p{4.23a}.
To finish with $P^i_a$, its on-shell expression in the WZ gauge
contains only one constant $O(8)$ vector
\be\label{4.16}
P^i_a={1\over 7!8!}\epsilon_{ab_1\dots b_7}\eta_{b_1}\ldots
\eta_{b_7}p^i, \;\;p^i={\rm const}.
\ee

The $(++)$ projection of $P^\mu_a$ is an auxiliary superfield.
This can be seen from the $\alpha=A$ projection of \p{4.10}
\be\label{4.18}
P^{++}_A - \gamma^i_{A\dot A}\gamma^k_{b\dot A}Y^kP^i_b = 0.
\ee
With the help of \p{3.33}, \p{4.15} and \p{4.16} this implies
\be\label{4.19}
P^{++}_a={1\over 7!8!}\epsilon_{ab_1\dots b_7}\eta_{b_1}\ldots
\eta_{b_7}p^{++},
\ee
where
\be\label{4.20}
p^{++}={1\over 2}{\dot x}^ip^i={1\over 4}({\dot x^i})^2p^{--}={\rm const}.
\ee

Combining  equations \p{4.5},\p{4.15}, \p{4.16}, \p{4.19} and \p{4.20} we
find
\be\label{4.21}
P^\mu_a={1\over 7!8!}\epsilon_{ab_1\dots b_7}\eta_{b_1}\ldots
\eta_{b_7}p^{\mu},
\ee
where $p^\mu$ is a constant $D=10$ vector (see \p{4.2a}) with the following
light-cone projections
\be\label{4.22}
p^\mu=(p^{--},\; p^i,\; p^{++})= p^{--}(1,\; {1\over 2}{\dot x}^i, \;
{1\over 4}({\dot x}^i)^2).
\ee
This is the momentum of the massless superparticle
\be\label{4.23}
p^\mu p_\mu=-p^{--}p^{++}+(p^i)^2=0.
\ee
Now we are in a position to justify the kinematical requirement
\p{2.12}. On the chart \p{3.16} of $S^8$ the condition
$p^\mu\neq 0$ is equivalent to
$p^{--}\neq 0$. Given this, from \p{4.15} one derives the on-shell
equations for the superparticle variables:
\be\label{4.23a}
{\ddot x}^i=0,\;\;\; \dot\theta_{\dot A}=0.
\ee
The conclusion is that  \p{2.12} is a non-degeneracy condition for
the Chern-Simons action \p{2.1}, similar to the condition $e\neq 0$ on
the einbein in the ordinary relativistic particle theory
\p{1.2}.

We mention that eq.\p{3.30} can now  be solved too:
\be\label{4.24}
X^{++} = x^{++}+i\eta_a\gamma^i_{a\dot A}\theta_{\dot A}
\dot x^i, \ \ \ \dot x^{++}={1\over 2}(\dot x^i)^2.
\ee
The final expressions for the other world-line superfields are
\be\label{4.25}
X^{--}=2\tau^{--}, \;\; X^i=x^i+ i\eta_a\gamma^i_{a\dot A}
\theta_{\dot A}, \;\;
\Theta_A=\eta_A, \;\; \Theta_{\dot A}=\theta_{\dot A}+
{1\over 2}\eta_a\gamma^i_{a\dot A}{\dot x}^i.
\ee
The superparticle velocity is given by \p{3.1a}
\be\label{4.26}
{\dot x}^\mu=({\dot x}^{--}, {\dot x}^i, {\dot x}^{++})=
(2, {\dot x}^i, {1\over 2}({\dot x}^i)^2).
\ee
Comparing \p{4.22} and \p{4.26} we find
\be\label{4.27}
p^\mu={1\over 2}p^{--}{\dot x}^\mu.
\ee
Note that the right-hand side of
\p{4.27} is proportional to $D_a\Theta\gamma^\mu D_a\Theta\vert_{\eta=0}$
when computed in the light-cone gauge. In fact, the relation
\be\label{4.27a}
p^\mu \sim D_a\Theta\gamma^\mu D_a\Theta
\ee
is gauge-independent. It can be derived from the gauge-invariant equation
\be\label{4.27b}
p^\mu(\gamma_\mu D_a\Theta)_\alpha = 0,
\ee
which is a consequnce of the equations of motion \p{3.2}, \p{4.1} and \p{4.2}.

So, we have solved all the equations of motion following from the action
\p{2.1}. The results should be compared with those from the Brink-Schwarz
action
\be\label{4.28}
S_{BS}=\int d\tau\left[p_\mu(\dot x^\mu-i\dot\theta\gamma^\mu\theta)
-{1\over 2}ep_\mu p^\mu\right].
\ee
The latter is invariant under world-line diffeomorphisms $\tau\rightarrow
\tau'(\tau)$ and under the $\kappa$ transformations
\be\label{4.29}
\delta p_\mu=0,\; \; \delta\theta^\alpha=
p_\mu(\gamma^\mu)^{\alpha\beta}\kappa_\beta,\;\;
\delta\dot x^\mu=-i\theta\gamma^\mu\delta\theta, \;\;\delta
e=-4i\dot\theta^\alpha\kappa_\alpha.
\ee
The equations of motion following from \p{4.28},
\be\label{4.30}
\dot p_\mu=0,\; p^2=0,\; \dot x^\mu - i\dot\theta\gamma^\mu\theta -ep^\mu=0,\;
p_\mu(\gamma^\mu\dot\theta)_\alpha=0,
\ee
can be easily solved in the reparametrization and $\kappa$-symmetry
gauges $x^{--}=2\tau$ and $\theta_A=0$. Thus one can identify  the
physical modes corresponding to the $N=8$ supersymmetric action
\p{2.1} and the Brink-Schwarz action \p{4.28}.

As we have pointed out earlier, in the action \p{2.1}
$\kappa$ symmetry is completely replaced by local world-line
$N=8$ supersymmetry. Now we can show the on-shell relation between
these two symmetries. Suppose that we do not fix the
superconformal group completely, like in \p{3.19}, \p{3.25}, but instead
keep  only the $N=8$ supersymmetry parameter local, $\epsilon_a =
\delta\eta_a\vert_{\eta=0}$. The $N=8$ supersymmetry transformation of
$\theta^\alpha = \Theta^\alpha\vert_{\eta=0}$ is given by
\be\label{1000}
\delta \theta^\alpha = \epsilon_a \psi_a^\alpha,
\ee
where $\psi_a^\alpha =
D_a \Theta^\alpha\vert_{\eta=0}$.  Let us substitute in \p{1000} the
field-dependent parameter $\epsilon_a = p^{--}
\psi_a^\alpha\kappa_\alpha(\tau)$ and make use of the following relations:
\be\label{1001}
\psi^\alpha_a\psi^\beta_a = {1\over 16}(\gamma_\mu)^{\alpha\beta}
\psi_a\gamma^\mu\psi_a,
\ee
\be\label{3.1a}
\dot x^\mu -i\dot\theta\gamma^\mu\theta=
{1\over 8}\psi_a\gamma^\mu \psi_a .
\ee
Eq. \p{1001} can be proved with the help of the gauge \p{3.19} and
eq. \p{3.21} (but it is valid in any gauge), eq. \p{3.1a} is obtained
from \p{3.1} by hitting it with $D_a$. Putting all this in \p{1000}
we find the $\kappa$ transformation of $\theta^\alpha$ \p{4.29}.
The conclusion is that $\kappa$ symmetry emerges as a result of an
almost complete
 gauge fixing of the $N=8$ superconformal group and a partial
use of the equations of motion.

\section{Coupling to a D=10 supergravity and Maxwell background}

The action \p{2.1} describes a free superparticle theory. Therefore it is
of interest to see how it can be coupled to background $D=10$ fields.
Introducing a supergravity background is straightforward. It is
sufficient to replace the flat D=10 superspace vielbeins in \p{2.1}
by curved ones, $E_M^{\;\;\hat A}(Z^N)$. Here $Z^M = (X^\mu,
\Theta^\alpha)$ and $\hat A = (\hat \mu, \hat \alpha)$ are the
tangent-space vector and spinor Lorentz indices. After that \p{2.1} becomes
\be
S_{SG} = i\int d\tau d^8\eta \; P_{a\hat \mu} D_aE^{\hat \mu} ,
\label{SG}\ee
where we used the notation $D_aE^{\hat A} \equiv (D_aZ^M) E_M^{\;\;\hat
A}$. Note that the Lagrange multiplier is now  a tangent-space
Lorentz vector. The invariance of the action \p{SG} under $N=8$
world-line conformal supersymmetry, target-space diffeomorphisms and
tangent-space Lorentz transformations is manifest. As we saw earlier,
the consistency of the superparticle action crucially depended on the
abelian gauge invariance \p{2.10} of the Lagrange multiplier. This
invariance is not automatic in \p{SG}, and we have to make sure it
still works. The obvious generalization of \p{2.10} to the curved
case is
\be\label{covxi}
\delta P^{\hat \mu}_a = D_b\left[\xi_{abc}^{\hat \alpha} (\gamma^{\hat
\mu})_{\hat\alpha\hat\beta} D_cE^{\hat\beta}\right],
\ee
It is not hard to obtain the variation of \p{SG} with respect to \p{covxi}:
\be\label{var}
\delta S_{SG} =
 {i\over 2}\int d\tau d^8\eta \; (\xi_{abc}\gamma_{\hat \mu})_{\hat \alpha}
D_cE^{\hat \alpha} \left[D_aE^{\hat \beta}D_bE^{\hat \gamma} T_{\hat
\gamma\hat \beta}^{\hat \mu} + 2D_aE^{\hat B}D_bE^{\hat\nu} T_{\hat\nu\hat
B}^{\hat\mu}\right].
\ee
Here $T_{\hat A\hat B}^{\hat C}$ is the background supergarvity torsion.
It is clear that the second term in \p{var} can be compensated for by
the following variation of the Lagrange multiplier:
\be
\delta P_{b\hat \nu} = -(\xi_{abc}\gamma_{\hat \mu}D_cE)D_aE^{\hat B}
T_{\hat \nu\hat B}^{\hat \mu}.
\ee
As to the first term in \p{var}, it vanishes due to the
$\gamma$-matrix identity \p{id} and the symmetry of $\xi_{abc}$,
provided we impose the following $D=10$ supergravity torsion
constraint \cite{N}:
\be\label{cons}
T_{\hat
\alpha\hat \beta}^{\hat \mu} = (\gamma^{\hat \mu})_{\hat
\alpha\hat \beta}.
\ee
The conclusion is that the consistency of the superparticle action
requires constraints on the background. This phenomenon is well-known
\cite{W}. In the Brink-Schwarz action one demands compatibility of the
background with $\kappa$ symmetry, and that leads to the constraint
\p{cons}. In our case $\kappa$ symmetry is replaced by world-line
conformal supersymmetry, which is manifest in \p{SG}. Instead, we had
to make the background compatible with the Lagrange multiplier gauge
invariance \p{covxi}, which lead us to the same constraint.

The next issue we would like to discuss is the coupling of the superparticle to
a Maxwell supersymmetric background. This can be done in the
framework of a
Kaluza-Klein scenario. Namely, we extend the action \p{SG} by adding
two new terms:
\be\label{5.1}
S_{SG+M}=i\int d\tau d^8\eta  \; \left(P_{a\hat \mu} D_aE^{\hat \mu}+ P_a
D_aA + P_aD_aX \right).
\ee
Here $D_aA \equiv D_aZ^M A_M(Z)$, where $A_M = (A_\mu, A_\alpha)$ are
the Maxwell background connections defined modulo abelian gauge
transformations:
\be\label{5.2}
\delta A_M=\partial_M\lambda(Z).
\ee
We emphasize the absence of a Maxwell coupling constant (electric
charge) in the action \p{5.1}. Instead, there we find a new
world-line superfield $P_a(\tau,\eta)$, which plays the same r\^ole
of a Lagrange multiplier for the Maxwell term
as $P_{a\hat \mu}$ plays for the first term.
In what follows we shall show that
on shell the only surviving component of $P_a$ is just the electric
charge of the superparticle. The superfiled $X(\tau,\eta)$
can be interpreted as an additional Kaluza-Klein bosonic coordinate of
target superspace.  Indeed, the action \p{5.1} is invariant under the
Maxwell transformations \p{5.2} provided
\be\label{5.3}
\delta P_a=0,\;\; \delta X=-\lambda.
\ee
Then the Maxwell superfield $A_M$ can be treated as part of the
supervielbeins $E_M^{\;\;\hat A}$ in a D=11 target space, according to
the standard Kaluza-Klein philosophy. The action is also
world-line superconformally invariant if we take $X$ to be a
scalar, and $P_a$ to transform in the same way as $P_{a\mu}$ in
\p{2.9}.

As in the case of supergravity, we have to make sure that the Maxwell
terms in \p{5.1} are invariant under
the following abelian gauge transformations of the Lagrange multipler $P_a$
(cf. \p{4.3}):
\be\label{5.4}
\delta P_a=D_b\xi_{ab},
\ee
where $\xi_{ab}(\tau,\eta)$ is an arbitrary symmetric and traceless
parameter. The variation of \p{5.1} is easily shown to be
\be\label{maxi}
\delta S_{SG+M}
={i\over 2}\int d\tau d^8\eta \; \xi_{ab} \left(D_aE^{\hat \alpha}D_bE^{\hat
\beta} F_{\hat\beta\hat\alpha} + 2 D_aE^{\hat A} D_bE^{\hat \nu}
F_{\hat \nu \hat A}\right),
\ee
where $F_{\hat A\hat B}$ is the Maxwell field-strength.
Once again, the second term in \p{maxi} can be compensated for by
\be
\delta P_{b\hat \nu} = -\xi_{ab}D_aE^{\hat A} F_{\hat \nu\hat A},
\ee
and the first term vanishes after imposing the $D=10$ Maxwell
constraint \cite{W}
\be\label{maxc}
F_{\hat\alpha\hat\beta} = 0.
\ee
Note that \p{maxc} is at the same time the integrability condition
for the equation obtained from varying \p{5.1} w.r.t. $P_a$:
\be
D_aX = -D_a A.
\label{integr}
\ee
In other words, due to \p{maxc} eq. \p{integr} allows to solve for
$X(\tau,\eta)$ in terms of the other fields (up to a constant),
without imposing any new restrictions on them.

Now we come to the important point about the origin of the electric
charge in the action
\p{5.1}. Varying it with respect to the Kaluza-Klein field $X$, one
gets $D_aP_a=0$. Repeating the arguments of section 4, we see that
this equation, together with the gauge invariance \p{5.4} leave just
one constant component in $P_a$ (cf. \p{4.3} and \p{4.5}):
\be\label{5.8}
P_a={1\over 7!8!}\epsilon_{ab_1\dots b_7}\eta_{b_1}\ldots
\eta_{b_7}e, \;\;e={\rm const}.
\ee
Substituting this into \p{5.1} gives, in particular, the following
component term
\be\label{5.9}
\int d\tau e \dot x^\mu A_\mu(x),
\ee
which is the usual Maxwell coupling for a charged particle.
Thus, we conclude that the integration constant $e$ is indeed the
electric charge of the superparticle.\footnote{The origin of the
electric charge and the string tension as integration constants was
recently discussed in two interesting papers \cite{T}.}

We would like to point out that the existence of the Maxwell coupling
\p{5.1}, which is invaraint under the $N=8$ superconformal group
and under $D=10$ superdiffeomorphisms, is a highly
nontrivial phenomenon. Suppose that we were given the electric charge $e$,
the Maxwell superfields $A_\mu, A_\alpha$, the
superparticle coordinates $X^\mu(\tau,\eta),\Theta^\alpha(\tau,\eta)$
and their derivatives. Then by simple dimensional arguments
 we would immediately conclude that it is impossible to construct any
off-shell $N=8$ invariant coupling {\it bilinear} in the Maxwell
fields and the target-space coordinates (as required by \p{5.9}).
The striking property of the Maxwell coupling \p{5.1} is the absense of an
off-shell electric charge. Instead, it is {\it trilinear} in the fields,
and only {\it on shell}, where one of the fields (the Lagrange multiplier
$P_a$)  reduces to the constant $e$, the action becomes bilinear.

The mechanism explained above may suggest a loophole in the various
``no-go'' theorems (see, for instance, \cite{RS})
that forbid the existence of off-shell supersymmetric
actions for theories like $D=10, \;N=1$ or $D=4, \;N=4$ super-Yang-Mills, etc.
 The point is that those theorems always assume {\it bilinearity} in
the fields. As we have seen, this assumption could be wrong off
shell, although it should definitely hold on shell (or rather upon
elimination of the auxiliary fields).\footnote{The idea that manifest
Lorentz invariance may require non-linearity in the auxiliary field
sector was first put forward by Siegel \cite{SIEG}.}

\section{Complex structures and Grassmann analyticity in the cases
$N=2 \; D=4$ and $N=4 \; D=6$}

The analysis of the superparticle action \p{2.1} carried out in the
most complex case $N=8 \; D=10$ can easily be adapted to the simpler cases
of the superparticle moving in $D=3, 4$ and $6$ superspaces (where
the identity \p{id} holds). To this end one should consider
$\Theta^\alpha$ as {\it real} (Majorana) spinors in those dimensions, and
the world-line should have $N=1,2$ and $4$ conformal supersymmetry,
correspondingly.\footnote{Of course, one can consider other
combinations of $D$ and $N$ (see, for instance, \cite{TONIN}, where the
case $N=2 \; D=10$ is discussed). However, it is only when $N$
reaches its maximal allowed value $D-2$ that the $\kappa$
symmetry of the superparticle can be entirely replaced by world-line
conformal supersymmetry.}  As we already mentioned, the $N=1 \; D=3$ analog of
\p{2.1} is just the STVZ action \cite{STV}.  An essential feature of
all these actions is the absense of any complex structure: all
superfields and gauge parameters are {\it real } functios of {\it
real} variables.  However, the $N=2 \; D=4$ and $N=4 \; D=6$ cases
allow for a different treatment \cite{DS}, which utilizes the
concept of {\it complex} structures inherent in four and six
dimensions, and the related concept of Grassmann analyticity. In
this section we shall show how those two alternative treatments, the
real and the complex ones, are related to each other.

In four dimensions  the Majorana spinor $\Theta^\alpha$ has four real
components.  However, one can think of it as a complex two-component
Weyl spinor (and its conjugate). Indeed, using the matrix $\gamma^5$
 (which is an example of a complex structure, $(\gamma^5)^2=
-1$) one can construct projection operators that split
$\Theta^\alpha$ into two complex halves.  In other words, one can
introduce the well-known two-component spinor notation
\be\label{6.1}
(X^\mu, \ \Theta^\alpha) \ \ \rightarrow \ \ (X^{\alpha\dot\alpha},
\ \Theta^\alpha,\ \bar\Theta^{\dot\alpha}),
\ee
where now $\alpha$ and $\dot\alpha$ take two values. Note also that
the matrix $\gamma^5$  generates a $U(1)$ automorphism of the $D=4$
supersymmetry algebra, which {\it commutes} with the Lorentz group $SO(1,3)
\sim SL(2,C)$. In the two-component formalism eq.\p{3.1} has the form
\be\label{101}
D_aX^{\alpha\dot\alpha} - i(D_a\Theta^\alpha)\bar\Theta^{\dot\alpha}
 - i(D_a\bar\Theta^{\dot\alpha})\Theta^{\alpha} = 0.
\ee
Here $a = 1,2$ (we recall that in the case $D=4$  the world-line has $N=2$
supersymmetry). Further, eq.\p{3.2} now becomes:
\be\label{102}
D_a\Theta^\alpha D_b\bar\Theta^{\dot\alpha} + D_b\Theta^\alpha
D_a\bar\Theta^{\dot\alpha} = \delta_{ab} D_c\Theta^\alpha
D_c\bar\Theta^{\dot\alpha} .
\ee
Introducing the complex notation
\be\label{103}
D=D_1+iD_2, \ \ \ \bar D = D_1 - iD_2,
\ee
one can rewrite \p{102} as follows:
\be\label{104}
D\Theta^\alpha  D\bar\Theta^{\dot\alpha} = 0.
\ee
This equation has two possible solutions: $D\Theta^\alpha =0$ or
$D\bar\Theta^{\dot\alpha} = 0$. Both of them cannot vanish at the
same time, since this would contradict the basic non-degeneracy condition
\p{2.11}. Without loss of generality we can choose the solution
\be\label{105}
D\bar\Theta^{\dot\alpha} = 0 \ \ \rightarrow \ \ \bar D \Theta^\alpha =0.
\ee
These are nothing but chirality conditions for the superfields
$\Theta^\alpha$ and $\bar\Theta^{\dot\alpha}$. Thus we see that the
natural complex structure of the $D=4$ spinors (represented by
$\gamma^5$), together with the Lorentz-harmonic defining condition
\p{3.2} induce a complex structure in the world-line superspace of the
superparticle. Further, eq.\p{101} implies chirality of $X$ as
well. Indeed, introducing the notation
\be\label{106}
X^{\alpha\dot\alpha}_L = X^{\alpha\dot\alpha} +
i\Theta^\alpha\bar\Theta^{\dot\alpha}, \ \ \ X^{\alpha\dot\alpha}_R =
X^{\alpha\dot\alpha} -
i\Theta^\alpha\bar\Theta^{\dot\alpha},
\ee
we can rewrite \p{101} as a chirality condition:
\be\label{107}
\bar DX^{\alpha\dot\alpha}_L = 0 \ \ \mbox{or} \ \
DX^{\alpha\dot\alpha}_R = 0.
\ee
So, in the case $N=2 \; D=4$ the physical content of the superparticle
is encoded in the chirality conditions \p{105}, \p{107} and the
definition \p{106}.  Actually, as it was explained in \cite{DS}, one can
reverse the argument as follows. Off shell one does not impose
the relations \p{106}.
Solving the chirality conditions \p{105},
\p{107} in suitable chiral bases in the world-line superspace, one considers
$X_L^{\alpha\dot\alpha}$ and $\Theta^\alpha$ as {\it unconstrained} chiral
superfields. They become the basic dynamical variables.
Then one  treats eq.\p{106} as a definition of the real
coordinate $X^{\alpha\dot\alpha} =1/2(X^{\alpha\dot\alpha}_L+
X^{\alpha\dot\alpha}_R)$,  as well as an equation of motion,
\be\label{108}
{i\over 2}(X^{\alpha\dot\alpha}_L - X^{\alpha\dot\alpha}_R) +
\Theta^\alpha\bar\Theta^{\dot\alpha} = 0.
\ee
The latter can be obtained from the action
\be\label{109}
S =\int d\tau d^2\eta P_{\alpha\dot\alpha}
\left( {i\over 2}(X^{\alpha\dot\alpha}_L - X^{\alpha\dot\alpha}_R) +
\Theta^\alpha\bar\Theta^{\dot\alpha}\right).
\ee
This is an alternative form of the $N=2 \; D=4$ action. The essential
difference between the two actions is that \p{2.1} is substantially real,
while \p{109} makes use of the concept of chirality (holomorphicity)
inherent in this case. Note also that the Lagrange multiplier
$P_{\alpha\dot\alpha}$ in \p{109} has the same dimension as the
superparticle momentum and it does not possess any abelian gauge
invaraiance like \p{2.10}. The on-shell contents of these theories are,
however, identical.

The case $N=4 \; D=6$ can be treated in a similar way. There the Majorana
spinors have 8 real components. At the same time they can be considered as
four-component Weyl spinors with an extra $SU(2)$ doublet index and a
pseudoreality condition (see, for instance, \cite{KT}). This $SU(2)$
group is an automorphism group of the $D=6$ supersymmetry algebra which
commutes with the $D=6$ Lorentz group $SO(1,5) \sim SU^*(4)=SL(2,H)$.
Thus in this case there are three complex structures
(in other words, a quaternionic structure) corresponding to the three
generators of the $SU(2)$ group.  The world-line Grassmann coordinates
$\eta_a$ ($a=1,\ldots,4$) are now $SO(4) = SU(2)\times SU(2)$ spinors, so
they also have a natural $SU(2)$ structure. Using all this, one can
show that the basic superparticle equation \p{3.1} and its consequence
\p{3.2} lead to the concept of $SU(2)$ harmonic analiticity
\cite {GIKOS}. From that
one can derive the alternative $N=4 \; D=6$ superparticle action
proposed in \cite{DS}. Note, however, that the harmonic action of
\cite{DS} uses infinite sets of auxiliary fields (coming from the
harmonic expansions of the world-line superfields), whereas the new
action \p{2.1} involves only finite sets. Also, $SU(2)$ harmonic
analyticity appears in the new formulation on shell only, while in
the approach of \cite{DS} all the off-shell dynamical variables are by
definition analytic.
Again, the on-shell content of the real and the $SU(2)$-analytic formulations
coincide.

A remarkable feature of the action \p{2.1} is its
universality:  it equally well describes all the magic cases
$D=3,4,6,10$ with the corresponding maximal $N=1,2,4,8$ world-line
supersymmetry.
In the lower-dimensional cases $D=4$ and $6$
it reproduces the specific properties of analyticity on shell.
The latter reflect the existence of complex structures there,
which are in turn related to the  automorphisms of the supersymmetry
algebra in $D$ dimensions  commuting with the Lorentz group
$SO(1,D-1)$.  At the same time, the action \p{2.1} does not require
any analyticity/complex structures {\it off shell}.

In ten dimensions the real 16 component spinors are both Majorana and
Weyl. There there exists no automorphism of the
supersymmetry algebra $\{D_\alpha,D_\beta\} = \gamma^\mu_{\alpha\beta}
\partial_\mu$ that  commutes with the Lorentz group
(except for the irrelevant scale transformations). This probably
explains the failure of many attempts to generalize the notion of
complex structure to the case $N=8 \; D=10$ and to find ``octonionic
analyticity''. The action \p{2.1} escapes from this problem in a very simple
way: it does not refer to any complex structures at all.

\section{Conclusions}

In this paper we have presented a new formulation of the $D=10$ superparticle
theory. The new action \p{2.1} propagates the same modes  as the
Brink-Schwarz one \p{4.28}. At the same time they have essentially different
 symmetries. While the $\kappa$ symmetry of the BS action forms
an algebra only modulo the equations of motion, all the symmetries of
the new action are realized linearly and close off shell. In other
words, the problem of finding auxiliary fields for $D=10$ BS superparticle
was solved. The $\kappa$ symmetry of the BS action \p{4.28} can now
be understood as an on-shell and partially gauge-fixed form of conformal
world-line supersymmetry.

Note that in the new formulation the specific r\^ole of $\kappa$
symmetry in constraining the Yang-Mills or supergravity background
is played not by the superconformal group (which is manifest), but by
the abelian gauge group of the Lagrange multiplier \p{2.10}. The
latter, however, is not directly related to $\kappa$ symmetry. For
instance, in the $N=1 \;D=3$ case the right hand side of \p{2.10}
vanishes, while $\kappa$ symmetry is non-trivial.

It should be emphasized that in this paper we studied the classical theory
of the superparticle. After having understood the structure of its symmetries,
it should now be possible to attack the problem of  Lorentz-covariant
quantization of this theory.

In a future publication an $N=8 \; D=10$ superstring action of the
type \p{2.1} will be presented. The specific Wess-Zumino term in the
superstring action is in many respects analogous to the Maxwell
coupling of the superparticle, with a two-form abelian gauge field
instead of the Maxwell field. This term will involve a separate
Lagrange multiplier, which will produce  the string tension on shell
as an integration constant (like the electric charge in section 5;
see also \cite{T}).

Another direction of possible development is related to constructing
off-shell formulations of the $D=10$ super-Yang-Mills and supergravity
theories.  The non-linear structure of the Maxwell coupling \p{maxi}
may shed new light on this old problem.

\vskip15mm
{\bf Acknowledgements}
The authors would like to thank E.Bergshoeff, N.Berkovits, E.Ivanov
and V.I.Ogievetsky and S.Shnider for stimulating discussions and Prof. J.Ellis
for hospitality at the TH Division of CERN, where part of this work
has been done.

\end{document}